\documentclass[lettersize,journal]{IEEEtran}
\usepackage{amsmath,amsfonts}
\usepackage{algorithmic}
\usepackage{array}
\usepackage[caption=false,font=normalsize,labelfont=sf,textfont=sf]{subfig}
\usepackage{textcomp}
\usepackage{stfloats}
\usepackage{url}
\usepackage{verbatim}
\usepackage{graphicx}
\usepackage{comment}
\usepackage{graphicx}
\usepackage{array}
\usepackage{graphicx}
\usepackage{graphics}
\usepackage{pdfpages}
\usepackage{amsmath,amssymb,amsfonts}
\usepackage{algorithmic}
\usepackage{graphicx}
\usepackage{array}
\usepackage{graphics}
\usepackage{textcomp}
\usepackage{comment}
\usepackage{rotating}
\usepackage{adjustbox}
\usepackage{lscape}
\usepackage{hyperref}
\usepackage{amsmath}
\usepackage{multirow}
\usepackage{graphicx}
\usepackage{xcolor}
\usepackage[T1]{fontenc}
\usepackage[switch]{lineno}
\usepackage{color,soul}
\def\BibTeX{{\rm B\kern-.05em{\sc i\kern-.025em b}\kern-.08em
    T\kern-.1667em\lower.7ex\hbox{E}\kern-.125emX}}
\usepackage{balance}
\begin{document}
\title{Driver State Modeling through Latent Variable State Space Framework in the Wild}
\author{Arash~Tavakoli, Steven Boker, and Arsalan Heydarian
\IEEEcompsocitemizethanks{\IEEEcompsocthanksitem Arash Tavakoli and Arsalan Heydarian (Corresponding Author) are with the Department
of Engineering Systems and Environment, University of Virginia,
VA, 22901.\protect\\
E-mail: ah6rx@virginia.edu
\IEEEcompsocthanksitem Steven Boker is with the department of Psychology, University of Virginia,
VA, 22901.\protect
}}

\markboth{Driver State Modeling through Latent Variable State Space Framework in the Wild}%
{How to Use the IEEEtran \LaTeX \ Templates}

\maketitle

\begin{abstract}
Analyzing the impact of the environment on drivers’ stress level and workload is of high importance for designing human-centered driver-vehicle interaction systems and to ultimately help build a safer driving experience. However, driver’s state, including stress level and workload, are psychological constructs that cannot be measured on their own and should be estimated through sensor measurements such as psychophysiological measures. We propose using a latent-variable state-space modeling framework for driver state analysis. By using latent-variable state-space models, we model drivers' workload and stress levels as latent variables estimated through multimodal human sensing data, under the perturbations of the environment in a state-space format and in a holistic manner. Through using a case study of multimodal driving data collected from 11 participants, we first estimate the latent stress level and workload of drivers from their heart rate, gaze measures, and intensity of facial action units. We then show that external contextual elements such as the number of vehicles as a proxy for traffic density and secondary task demands may be associated with changes in driver's stress levels and workload. We also show that different drivers may be impacted differently by the aforementioned perturbations. We found out that drivers’ latent states at previous timesteps are highly associated with their current states. Additionally, we discuss the utility of state-space models in analyzing the possible lag between the two constructs of stress level and workload, which might be indicative of information transmission between the different parts of the driver’s psychophysiology in the wild.   
\end{abstract}


\section{Introduction}
Understanding driver's state, including stress level, emotions, and cognitive load, is one of the important factors to improve driver-vehicle interactions and to enhance driver safety, comfort, and experience \cite{bustos2021predicting,tavakoli2021harmony,jeon2017emotions,du2020psychophysiological}. Recent studies are pointing towards the effect of drivers' mental state on drivers' performance both in manual and automated driving \cite{agrawal2021evaluating}. For instance, drivers' stress levels and negative emotions were shown to be associated with higher accident rates \cite{dingus2016driver}. Similarly, cognitive load due to distraction such as secondary task engagements has an adverse effect on drivers' take-over ability in automated driving and is associated with higher accident rates \cite{odachowska2021psychological,dingus2016driver}. In the definition of the National Highway Traffic Safety Administration (NHTSA) of automation (from level 1, driver assistance, to 5, fully automation), both levels 2 and 3 require driver take-over control in immediate and with notice conditions, respectively. A safe vehicle at levels 2 and 3 is reliant on drivers' attention to take over in a timely fashion, which is also reliant on drivers' state. Thus, a proper take-over requires appropriate detection, modeling, and analysis of drivers' states in different driving scenarios.       

Driver's states, including emotions, cognitive load, and stress level, are psychological constructs that cannot be measured directly and are often observed through a set of measurement variables such as psychophysiological records of the driver, including driver's heart rate, gaze variability, brain signals, and skin conductance. One method to model psychological constructs such as stress level, workload, and emotions is to use latent variable modeling schemes. A latent variable framework models one or multiple unobserved constructs through a set of sensory measurements \cite{hunter2018state}. Each sensor measurement might reveal a part of the real construct with a certain level of error. For example, the stress level might be modeled through a set of physiological measures such as heart rate (HR), skin temperature, and skin conductance through latent variable modeling \cite{kim2018stress}. Similarly, cognitive load can be measured through driver's gaze variability, brain activity, and heart rate measurements \cite{lohani2019review,engstrom2017effects}. 

Additionally, considering that psychological constructs in driving often happen simultaneously, it is thus required to analyze the driver's state in a holistic fashion where different constructs exist simultaneously and can interact with each other. For example, real-world driving, which can be accompanied by stress-inducing driving events, often happens together with cognitive tasks such as the task of driving itself as well as the secondary tasks \cite{lohani2019review,tavakoli2021driver}. The real-world analysis of driving situations often has many, if not all of the psychological constructs interacting with each other in a dynamic fashion. In a hypothetical driving scenario, a driver might become frustrated by a lead vehicle's sudden stop; this might increase the driver's stress level due to fear of hitting the lead vehicle, while the driver might also be working with a phone or mind wandering. A review on the interaction between affect and driving behaviors also points out that it is required for driving behavior research to integrate affect, cognition, and behaviors in one framework for a more holistic understanding of the interactions among these elements \cite{jeon2015towards}. Latent variable modeling can thus become helpful in modeling not just the psychological constructs of interest but also their possible interaction over time.

While latent variable modeling of psychological constructs helps with realistic modeling of human's state, modeling driver's state should also address the problem of the time dependency of driver state measurements \cite{sadeghi2021posttraumatic}. Traditionally, when modeling time-dependent sensory measurements such as human HR, modeling techniques such as autoregressive moving average are used where the dependency of time points can be modeled explicitly. The problem of time dependency is also present when analyzing latent constructs where the time dependency of latent variables needs to be evaluated in the modeling scheme. For instance, when modeling stress level as a latent variable based on physiological sensors, it is important to understand the changes in stress level during the driving scenario and to what degree the stress level at each time point is related to the next. This can be achieved through state-space modeling \cite{hunter2018state}. In other words, state-space modeling can be imagined as a substitute for autoregressive models in a latent variable framework. 

Lastly, driving happens in a multidimensional \textit{contextual setting}. In short \textit{context} can be defined as any information that helps to define a situation in driving that is either internal or external to the driver. External context includes sensor measurements from the environment (e.g., weather condition, temperature, and traffic density), as well as the vehicle (e.g., current speed). The internal context consists of the measurements related to the driver (e.g., stress, valence, and arousal levels) \cite{tavakoli2021harmony}. Different parts of the context can interact and affect each other. Recent studies in psychology are emphasizing the importance of context in human state analysis in which emotional episodes are tied to situation-specific needs of humans \cite{ortony1990cognitive,hoemann2020context,barrett2011context}. In other words, while analyzing latent human constructs, it is important to account for the effect of external contextual inputs explicitly to achieve a more realistic model. To achieve these aforementioned modeling schemes, we propose using state-space latent variable models for driver state analysis, where the effect of external context can be measured explicitly on the latent constructs and in a temporal fashion.

The rest of the paper is as follows. We first provide detailed background literature on driver's psychophysiological states (i.e., workload and stress levels) and how they affect driver sensing data (e.g., HR), as well as their interaction with the external context (e.g., traffic density). We then propose a framework for analyzing driver's psychophysiological measures together with the external context longitudinally through taking advantage of latent variable state-space models. We showcase a sample exploratory analysis of the state-space approach on real-world driving data. We build state-space models to analyze the relationship between contextual elements (i.e., number of road users on the road and driver's task demands) and a driver's stress and workload as latent variables estimated through driver's facial expressions, eye gaze patterns, and physiological responses. Through our case study, we compare two path diagram models in a state-space modeling scheme by considering (1) one latent variable for drivers' psychophysiological state, (2) two separate latent variables for driver's psychophysiological state with an interaction with each other. We assess the log-likelihood of the aforementioned models as a measure of model fitness and discuss the comparison. We then analyze the time dependency of the aforementioned latent variables. This research takes a holistic approach to driver state analysis where multiple internal and external factors are unified through one modeling framework temporally. Through our case study, we specifically answer the following questions:

\begin{enumerate}
    \item How do the external context (i.e., traffic density) and task demands (based on hand-movement)  affect the driver's internal contextual state (i.e., stress level and workload)? 
    \item How does the latent state at previous timestamps (i.e., 1 through 10 seconds) of the driver affect his/her current state (i.e., stress and workload)? 
    \item How does the association between latent constructs (stress level and workload) change throughout the driving scenario?
    
\end{enumerate}

\section{Background Study}
The background section is divided into three subsection of context (\ref{sec:context}) , stress level (\ref{stress}), and workload (\ref{sec:work load}) as follows. 

\subsection{Context} \label{sec:context}
Depending on the field of study, the context has been defined with different elements. In computing systems, \cite{schilit1994disseminating} defined it as surrounding objects, their locations, and their variation over time. A similar study, defined aspects of context to be \emph{who you are}, \emph{who you are with}, and \emph{what resources are nearby} \cite{schilit1994context}. Based on this study, context is not just the location but also noise level, lighting, social situation, etc. Later \cite{dey2001conceptual} defined context as any relevant information that can help characterize the situation of the interaction between an entity, an application, and the environment. Two keywords in this definition are \emph{any relevant information} and the \emph{interaction}. This study points out that the relevant information is not always location or objects, but it is defined based on the situation that an entity is in. More importantly, it defines certain types of context to be location, identity, activity, and time as primary. Secondary context can be derived from the primary ones. For instance, a person's identity can define their home address. Another study has defined context in three layers of adaptive layer, being responsible for retrieving the context from sensors, management layer being responsible for providing this information to other devices, and application layer being the part that uses underlying information for different applications \cite{hofer2003context}. The current study considers five types of context: time, location, device, user, and network. It also points out that contextual information can be physical or logical, in which the physical part refers to raw sensors readings (e.g., body temperature), and logical context is their abstract meanings (e.g., a person having a fever).  

The field of psychology provides a more human-centered definition for context. For instance, \cite{hofer2003context} points out the importance of considering emotions and preferences when defining context. However, it is recognizable that humans' emotions and actions by themselves are meaningful when considered in the external context that they are happening \cite{greenaway2018context}. This points towards separating context into two categories, internal and external, where internal context is related to the human, while external context is concerned with the environment. Other studies have also provided definitions for context in driving, mostly centered around the environment, which is categorized as part of the previously mentioned definitions (external context). In other words, environment and context seem to be interchangeably used in such studies. This includes road conditions, weather conditions, and anything that happens in the environment. \cite{schneemann2016context} defines context as road scene and segments. \cite{moosavi2017characterizing} defines context as a combination of environment and time. \cite{tivesten2014driving} defines context as the presence of turning maneuvers, a lead vehicle, and oncoming vehicles. 

In this paper, we define context similar to \cite{dey2001conceptual,fernandez2019contextual} as a collective summary of any relevant information about a driving situation retrieved from sensor measurements. However, by using the psychological theories definition for context, we note that context can be internal or external to the human driver. Internal contextual cues are retrieved from the human driver and can affect the driving situation (e.g., driver's heartbeat demonstrating calm or stressful state in a lane change action), whereas external contextual cues are related to the environmental elements affecting the driving situation (e.g., rainy road condition on a lane change situation).

\subsection{Stress} \label{stress}
Stress is the process in which the demand of a particular situation is perceived to be more than the available resources \cite{francis2018embodied}. The perceived demand can be defined based on the overall situation, including the previous experiences, internal body sensations, and the external stimuli \cite{francis2018embodied}. Stress can be in different time scales where short-time is referred to as acute versus long term, referred to as chronic \cite{francis2018embodied}. 

Decreasing drivers' stress is of high importance as it can contribute to human error making and possible accidents \cite{krahnen2022evaluation}. Studies in the past have pointed out the effect of stress level on drivers' performance and driving behavior \cite{matthews1998driver,ge2014effect}. For instance, \cite{ge2014effect} showed that perceived stress might affect behaviors such as aggressive driving and drunk driving. Multiple biomarkers have been used in literature for detecting stress \cite{balters2021individualized,chesnut2021stress}. Studies have pointed out the utility of cardiovascular measures in detecting the human state. Overall, cardiovascular activity can be measured through two technologies of Electrocardiography
(ECG) and photoplethysmogram (PPG). ECG measures the electrical activity of the heart through the usage of contact electrodes, whereas PPG records the same activity through measuring blood volume in the vein using infrared technology \cite{lohani2019review,tavakoli2021harmony,tavakoli2021leveraging}. Devices such as conventional wearable technologies often use PPG. The electrical signal of the heart can then be used to measure the HR (i.e., beeps per minute) and to calculate the heart rate variability (HRV) features. HRV is generally referred to as a set of features that are retrieved from the sequence of beat-to-beat intervals of an individual's HR signal. These features can be calculated in time (e.g., root mean squared of successive intervals (RMSSD)), frequency (e.g., low-frequency power (LF)), and non-linear domains (e.g., sample entropy). There is a vast literature showing that such features are, in fact, correlated with human stress levels \cite{kim2018stress}. An increase in stress level is shown to be correlated with a decrease in RMSSD, an increase in LF, and an increase in HR \cite{kim2018stress}. Additionally, studies pointed out the significance of individual differences in the relationship between subject stress level and HR \cite{sommerfeldt2019individual}. 

In addition to HR, certain facial action units were used to infer stress levels. For instance, \cite{gavrilescu2019predicting} reported that AU1, AU6, AU12, AU15 are the most indicative of stress levels in their study. However, when comparing dependent and independent subject methods when performing automatic facial emotion detection, the same study reported that the accuracy of stress detection using AUs dropped from 91 \% to 74 \% \cite{gavrilescu2019predicting}. Such results may show that individual differences exist in how people react in different situations. In other words, if the situation defines how a person might react, mere analysis of biological responses may not provide the true human state at each time point. These findings may suggest that driver's state analysis through psychophysiological metrics should take into account the specific situation and consider individual differences.

\subsection{workload} \label{sec:work load}
Driver's workload is mostly defined as cognitive resources that are taken from the driver by any activity other than the driving \cite{engstrom2017effects}, although some studies have also examined the workload from the driving itself \cite{mehler2019demanding}. The workload includes both ''mind wandering" and the load imposed by ''secondary task". Cognitive workload has been shown to affect driving performance metrics. Engstrom et al. point out that cognitive workload can selectively impair driving tasks that rely on cognitive control as opposed to automated tasks \cite{engstrom2017effects}. Studies have shown that cognitive workload might impair driver's object detection response \cite{horrey2006examining}, especially for the objects that are novel or difficult to detect \cite{engstrom2017effects}. Additionally, studies show that drivers' decision-making is also negatively affected by cognitive workload \cite{engstrom2017effects,cooper2003impact}. This highlights the importance of detecting and possibly mitigating drivers' cognitive workload.     

Multiple biological signals such as driver's eye metrics, cardiovascular measures, and brain signals have been used extensively for workload estimation in both in-lab and real-world situations \cite{lohani2019review,tavakoli2019multimodal,tavakolipersonalized}. Driver's eye metrics in these studies included patterns of blinking rates, saccades, fixations, stationary and transition entropy \cite{krejtz2018eye,fabio2015influence,shiferaw2018stationary,shiferaw2019review}. Here we direct our focus on a more recent feature of the driver's gaze, which is the driver's gaze entropy metric. There are two measures of entropy for a random variable. Information Entropy refers to the uncertainty associated with a choice \cite{shiferaw2019review}. The entropy increases with the higher levels of uncertainty in randomness in a system. This is generally calculated through Shannon's equation \cite{shannon1948mathematical}. In gaze analysis, this entropy refers to the overall predictability of fixation locations which is a measure of gaze dispersion \cite{shannon1948mathematical}, and is called Stationary Gaze Entropy (SGE). For a set of fixation locations in a sequence of eye movements, one can assign fixation locations to spatial bins of $p_i$ and calculate the SGE as:
\begin{equation}
SGE = -\sum_{\textit{i=1}}^{n} p_{i} \log_{2}p_{i} 
\end{equation}

SGE is used extensively in the literature for human state analysis. For instance, studies have pointed out the utility of SGE for detecting task demand, complexity, experience, workload, drowsiness, and being under the influence of alcohol \cite{shiferaw2019review}. Because inferences based on SGE can be very task-related, studies have moved more towards a new measure of entropy, namely conditional entropy. In other words, when assessing cognitive load during a task using SGE, it is important to know whether the task requires high or low SGE for optimum performance.

Conditional entropy takes into account the dependency between different fixations in a temporal fashion. This results in Gaze Transition Entropy (GTE). GTE is a measure of predictability of the next fixation location given the current location. For a sequence of transitions between different spatial bins of $i$ and $j$, with a probability of $p_{ij}$ the GTE is calculated as:
\begin{equation}
GTE = -\sum_{\textit{i=1}}^{n} p_{i} \sum_{\textit{j=1}}^{n} p_{ij} \log_{2}p_{ij} 
\end{equation}

Driver's GTE is shown to be correlated with higher task demand, higher scene complexity, and higher levels of workload \cite{shiferaw2019review,fabio2015influence}. Additionally, a recent theory suggests that conceptually for each specific combination of task demand and scene complexity, there exists an optimal GTE \cite{shiferaw2019review}. The optimal GTE is the result of an interaction between a human's internal state (e.g., memory) and the level of external information provided to the human through sensory inputs through the prediction process of the outside world. Deviating with an increased level of GTE (relative to the optimum) can be due to stress, anxiety, and emotional episodes, while a decrease in the level of GTE (relative to optimum) can be due to usage of depressants such as alcohol. 

Workload detection has also been done based on facial action units. For instance, \cite{yuce2016action} reported the top four correlated action units with workload detection are AU1, AU2, AU07, AU25. However, similar to stress level detection, individual differences in workload estimation were shown to be important. For instance, \cite{yuce2016action} showed that when performing subject-independent tests for workload estimation using facial AUs as compared to subject dependent, the accuracy dropped from 95 \% to 68 \%, an indication of inter-individual differences in facial reactions to workload. 

While multiple studies have advanced our knowledge on workload or stress separately in driving, not that many studies considered analyzing both constructs simultaneously. Studies in the past have provided evidence that certain human physiological measures may be impacted differently in the case of having both stress and cognitive load present in the situation. For example, a recent review \cite{shiferaw2018stationary} reports that GTE can increase from its optimal value for a specified task if states such as anxiety and stress are present while performing a task. Thus it is important to analyze drivers' cognitive load and stress level simultaneously and in a unified framework.



\section{Methodology}
In this section we outline details regarding the mathematics behind state space models (\ref{sec:state_space}), data collection (\ref{sec:data_col}), feature extraction (\ref{sec:feat}), analysis environment (\ref{sec:an_env}), and model selection (\ref{sec:mod_selection}).   
\subsection{A Latent Variable State-Space Model for Driver's State}  \label{sec:state_space}
One method to analyze driver's state is to consider the system of driver, vehicle, and the environment as a dynamical system in a state-space fashion where different perturbations from the environment change driver's state momentarily \cite{mirman2019dynamical}. A state-space model (SSM) is a mathematical representation for a dynamical system consisting of a set of inputs (referred to as perturbations) and outputs which are properties of the system that evolve over time and are measurements of certain latent variables that cannot be measured on their own \cite{lodewyckx2011hierarchical}. Using the observed variables, the latent state is estimated with a certain error, and the task of SSM is to provide the latent estimation as well as the effect of perturbations on those latent states \cite{lodewyckx2011hierarchical}. This is especially important for psychological constructs that are often measured through a set of observed variables, such as detecting stress (latent variable) through changes in physiological measures such as heart rate (observed variable). 

Another important benefit of using state-space models is their natural solution to the problem of time-dependent variables. First-order State-space models (SSM) analyze the system in a recursive manner where each time point is modeled based on the previous time point in a one-step Markov process fashion \cite{hunter2018state}. For the purpose of this article, we only consider first order (lag 1) state-space models. In this way, state-space models handle the time dependency of observations that is often the problem when using high-frequency sensor measurements (e.g., HR data). SSMs are suitable for analyzing driver's state since (1) driver's state is a latent variable (e.g., stress, workload, and emotion) measured through observed variables (e.g., driver's gaze, heart rate, and skin temperature); and (2) it is time-dependent in that events in the past can affect how a driver might feel and act in the future. 

\begin{figure}[ht]
\begin{center}
  \includegraphics[width=1\linewidth]{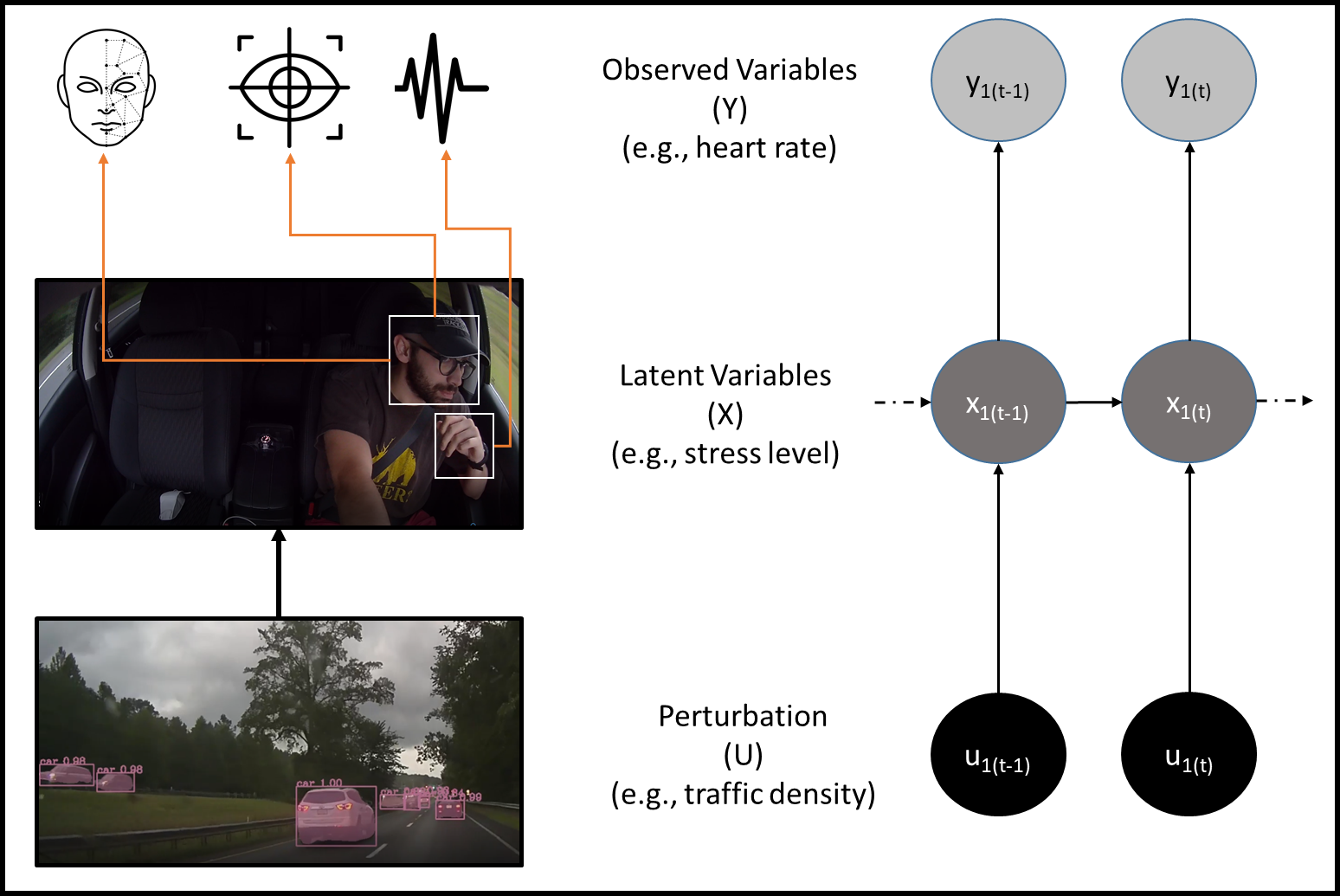}
  \end{center}
  \caption{A conceptual framework of a state space model consisting of perturbations, latent and observed variables at subsequent timesteps of (t-1) and (t).}
  \label{fig:frame}
\end{figure}

We provide a summary of the procedure behind SSM. The reader is referred to \cite{hunter2018state} and \cite{lodewyckx2011hierarchical} for greater details. Based on the notation provided in \cite{hunter2018state}, the general equations for a state-space model consist of two main equations of state equation \ref{state_eq}, and measurement equation \ref{measurement_eq}, as follows:

\begin{equation}\label{state_eq}
    x_t = Ax_{t-1} + BU_{t} + q_t
\end{equation}
\begin{equation}\label{measurement_eq}
    y_t = CX_t + DU_t + r_t
\end{equation}

where \textit{x} is the vector of latent variables at different time points, \textit{U} is the vector of perturbations to the system (i.e., inputs), \textit{$q_{t}$} is the vector of dynamic noise with covariance of \textit{Q}, \textit{$r_t$} is the vector of observation noise with zero mean and covariance of \textit{R}, and \textit{$y_t$} is the observations. Additionally, \textit{A} is the matrix defining the autoregressive components across time for latent variables, \textit{B} is the matrix measuring the effect of perturbation on the latent variables, \textit{C} is factor loading of each latent variable based on the observed variable, and \textit{D} is the matrix measuring the effect of perturbations on the observed variables. A schematic graph of this SSM representation is presented on Fig. \ref{fig:frame}.

As mentioned before, SSM analyzes the data in a recursive manner. SSM uses Kalman filter, which alternates prediction and correction steps as follows \cite{hunter2018state}. In summary, SSM works in two steps: first, it predicts the latent variable using initial values and the state model. Second, it uses the measurement model and observation variables to update the prediction. Using a latent variable matrix at each timestep, together with its covariance matrix (P), SSM predicts the latent variables at the next timestep as: 

\begin{equation}
    x_{t|t-1} = Ax_{t-1|t-1} + Bu_t
\end{equation}
\begin{equation}
    P_{t|t-1} = AP_{t-1|t-1}A^T + Q
\end{equation}

The prediction is then updated using the observed variables and the measurement model. In more detail:

\begin{equation}
    \hat{y_t} = Cx_{t|t-1} + Du_t
\end{equation}

The error is then calculated as:

\begin{equation}
    \tilde{y} = y_t - \hat{y_t}
\end{equation}

\begin{equation}
    \hat{S_t} = CP_{t|t-1}C^T + R
\end{equation}

\begin{equation}
    K = P_{t|t-1}C^T\hat{S}_t^{-1}
\end{equation}

\begin{equation}
    x_{t|t} = x_{t|t-1} + k\tilde{y}_t
\end{equation}

\begin{equation}
    P_{t|t} = P_{t|t-1} - KCP_{t|t-1}
\end{equation}

And lastly, the log likelihood (LL) of an SSM model is calculated as below \cite{hunter2018state}:

\begin{equation}
    -2LL = nlog(2\pi) + log|\hat{S}_t| + (y_{t} - \hat{y}_t)^T\hat{S}_t^{-1}(y_{t} - \hat{y}_t)
\end{equation}

where n is the number of observation variables.

\subsection{Data Collection} \label{sec:data_col}

The data for this study is provided by HARMONY, a human-centered multimodal driving study in the wild \cite{tavakoli2021harmony}. HARMONY is a framework that collects naturalistic longitudinal driving data through cameras, smart wearables, and multiple APIs. Through HARMONY, the driver's HR, hand movements (i.e., IMU sensors), facial expressions, gaze direction, pose direction, vehicle's speed, and location, as well as outdoor environmental videos, are collected automatically. We first focus on a long-term driving of one participant (\#9) on a highway for the purpose of this paper. Also, the data from this participant is available online for further research \cite{Harmonydata}. A snapshot of the driving scenario is depicted on Fig. \ref{fig:vis}. 

We then extend our findings across participants to assess the individual differences. To do so, we chose a random subset of the Harmony data from 10 other participants that drove in a long-term driving scenario (more than 1.5 hours) on a highway that was visually similar to that of the first participant. Note that because of the naturalistic nature of the data (which is not an on-road controlled study), participants drive on different roads. However, we only chose driving scenarios in highways of the state of Virginia to increase the similarity of driving scenarios across participants. More specifically, the data for this paper for each participant includes: 

\begin{itemize}
    \item \textbf{Smartwatch}: one channel of PPG signal sampled at 100 Hz, HR (HR) sampled at 1 Hz, and driver's hand acceleration and rotational velocity in three directions (i.e., X, Y, and Z) sampled at 100 Hz. 
    \item \textbf{Camera}: in-cabin and outside videos sampled at 30 Hz, with 1080p quality.
\end{itemize}

\subsection{Feature Extraction} \label{sec:feat}
\subsubsection{Video}
Using in-cabin videos, we retrieve the driver's gaze direction and facial AUs through applying OpenFace \cite{baltrusaitis2018openface} on the videos. OpenFace provides gaze angles in both X (horizontal), and Y (vertical) directions. These values are measured in radian, independent of the participant's head location. 

\textbf{Gaze Transition Entropy:} For this variable, based on method in \cite{krejtz2015gaze}, we first construct a 2D space of the range of the gaze angles in the study duration (i.e., 2-hour driving period). Then we divide the space into equally distanced areas of interest (AOI) as a 4*4 grid \cite{shiferaw2018stationary}. This results in a sequence of AOIs for the driver, which is related to different areas of the frontal view (e.g., front road, center stack, left side, and etc.). We then use the method provided in \cite{krejtz2015gaze} to calculate gaze transition entropy (GTE) for each time window in the driving scenario. In summary, for a sequence of AOIs, we first find the transition matrix between areas of interest by assuming a first-order Markov process for the gaze sequences \cite{krejtz2015gaze}. To this end, a transition matrix is retrieved with $p_{ij}$ being the probability of switching between AOIs i and j (in S), with a stationary probability of $\pi_i$. Then the GTE is calculated based on Shannon's entropy as:

\begin{equation}
\hat{H}_{t} = -\sum_{\textit{i} \in S} \pi_{i} \sum_{\textit{j} \in S} p_{ij} \log_{2}p_{ij} 
\end{equation}

\textbf{Facial Action Units:} We limit the facial AUs in our models based on previous literature to the ones that were shown to be correlated with stress levels and workload. As mentioned in background literature, \cite{gavrilescu2019predicting} reported that AU1, AU6, AU12, AU15 are the most indicative of stress levels in their study. \cite{yuce2016action} reported the top four correlated action units with workload detection to be AU1, AU2, AU07, AU25. Note that any other AU can also be used for the purpose of stress and workload estimation. We intentionally keep the number of AUs limited for a better model interpretation. 

\begin{figure}[ht]
\begin{center}
  \includegraphics[width=1\linewidth]{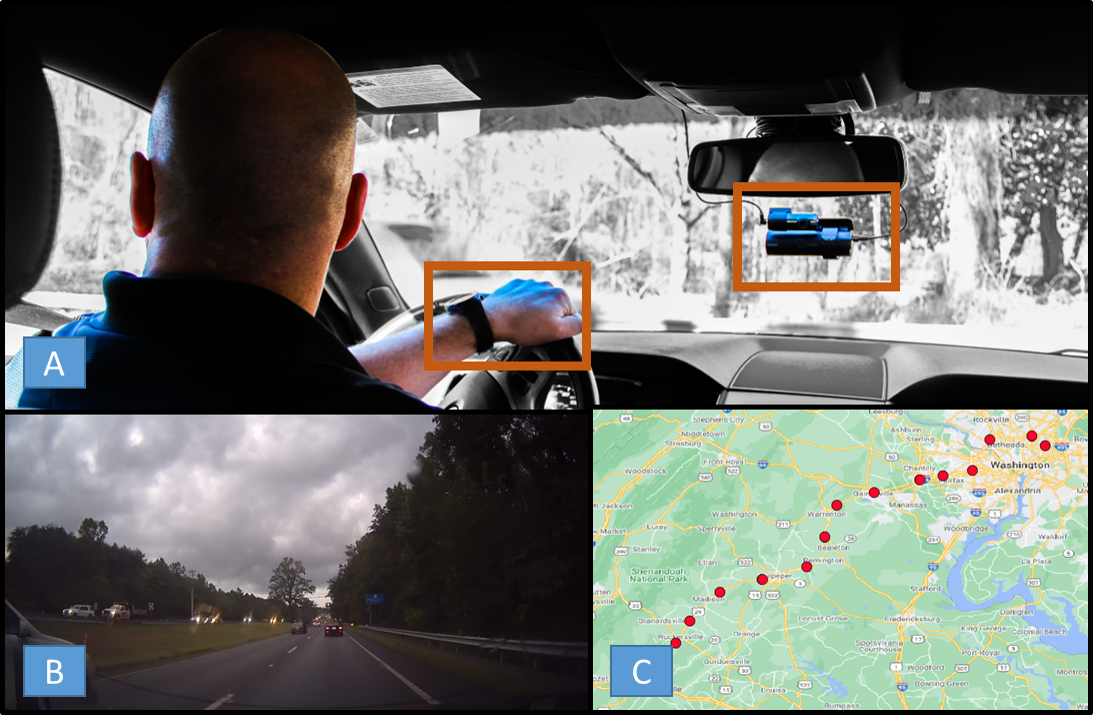}
  \end{center}
  \caption{A snapshot of the data used in this study including the devices (camera and smart wearable) (A), the road view (B), and the map of the case study trip (C)}
  \label{fig:vis}
\end{figure}

\textbf{Road Object Detection:} By using the outside videos and by applying Mask-RCNN \cite{he2017mask} algorithm trained on COCO dataset \cite{lin2014microsoft} as implemented by \cite{matterport_maskrcnn_2017}, we retrieve the number of cars, buses, pedestrians, bikes, motorcycles, stop signs, traffic lights, and trucks in each frame of the video. Here we define the scene complexity as the total number of road users in the field of view, which is the sum of cars, buses, pedestrians, motorcycles, and trucks (Fig. \ref{fig:vis}). Note that due to the relatively short length of dataset per person, we have not considered the effect of momentary perturbations such as traffic lights, stop signs, etc. This will be addressed in future work with a longer dataset. 

\subsubsection{Smartwatch}
\textbf{Heart rate:} Previous studies have shown the correlation between a human's heart rate and stress levels as described in section \ref{stress}, in which higher heart rate values might be indicative of higher stress levels. Additionally, our previous studies showed that certain stressors on the road could increase a driver's HR from its baseline value, in which change points in HR time series can be used to detect such stressors \cite{tavakoli2021harmony,tavakoli2021leveraging,tavakoli2021multimodal}. While HR values were sampled at 1 Hz, the exact frequency of sampling often changes within 0.9-1.1 Hz due to hardware issues. To address this, we resampled the HR values at 1 Hz frequency. We also apply Bayesian Change Point (BCP) detection to the driver's HR. Without going into the details of this method, BCP detects the specific moments that a change occurs in the underlying distribution of the data based on the Bayesian change point model provided in Barry and Hartigan's book \cite{barry1993bayesian}. In summary, this model assumes that the mean of the input within different segments remains constant. Change points in driver's HR might be correlated with stressful outside events where the posterior mean of the HR increases as a response to the stimulus\cite{tavakoli2021harmony}. This is in line with previous studies that higher HR values might be associated with higher stress levels. To apply this method, we use the BCP package in R \cite{erdman2007bcp}. The details of the BCP procedure are also provided in our former article \cite{tavakoli2021harmony}. 

\textbf{IMU:} In order to find out the moments that the driver's hand had abrupt movements, depicting activities such as working with a phone, we use the hand IMU sensor and find the magnitude of each of the gyroscope and accelerometer sensors. We then only consider the values above the average value of each sensor as timepoints when movements are abnormal.   

Each row of the final data frame after feature extraction includes driver's HR and its features (i.e., probability of a change and the posterior mean detected by BCP), driver's gaze entropy, the intensity of specific facial AUs (a number between 1-5), traffic density (i.e., number of road users in the field of view) as a measure of scene complexity, and driver's hand movement as a proxy for driver's non-driving related (secondary) tasks. Driver's HR, facial AUs, and gaze features are the observation variables for the internal context latent variables (i.e., stress level and workload), while scene complexity and task demands are the external contextual inputs to the system. 


\subsection{Analysis Environment} \label{sec:an_env}
We use the MARSS package \cite{holmes2012marss} developed in R programming language for estimating different SSM matrices of A, B, C, and D as described in section \ref{sec:state_space}. We propose two different models, and we compare these models to our data. The comparison is performed using the log-likelihood of each model, where a higher value shows a better fit to the data \cite{holmes2012marss}. Finally, we discuss the implications of the best fit model. The analysis is performed with a 10 seconds timestep look back, meaning that each timestep is compared with 10 seconds in the past. While we could also compare every two consecutive data points, increasing the timesteps helps with a better interpretation of the model. Also, previous research in driver behavior analysis, especially in the vicinity of a crash or near-crash event, generally considers a time point of 6-20 seconds prior to the crash event for prevalence analysis of different factors (e.g., driver emotion and distractions) \cite{dingus2016driver}. Also, smaller numbers increase the high dependency between timepoints, which makes the model interpretation more difficult. As the timepoints were too close in time, this step lets us understand the effect of perturbations in greater detail by removing the high dependency among the data points that are close in time. Note that this step does not remove any data point and only compares each timepoint at (t) with the data point at t+10 by restructuring the data.

\begin{figure*}[ht]
\begin{center}
  \includegraphics[width=0.95\linewidth]{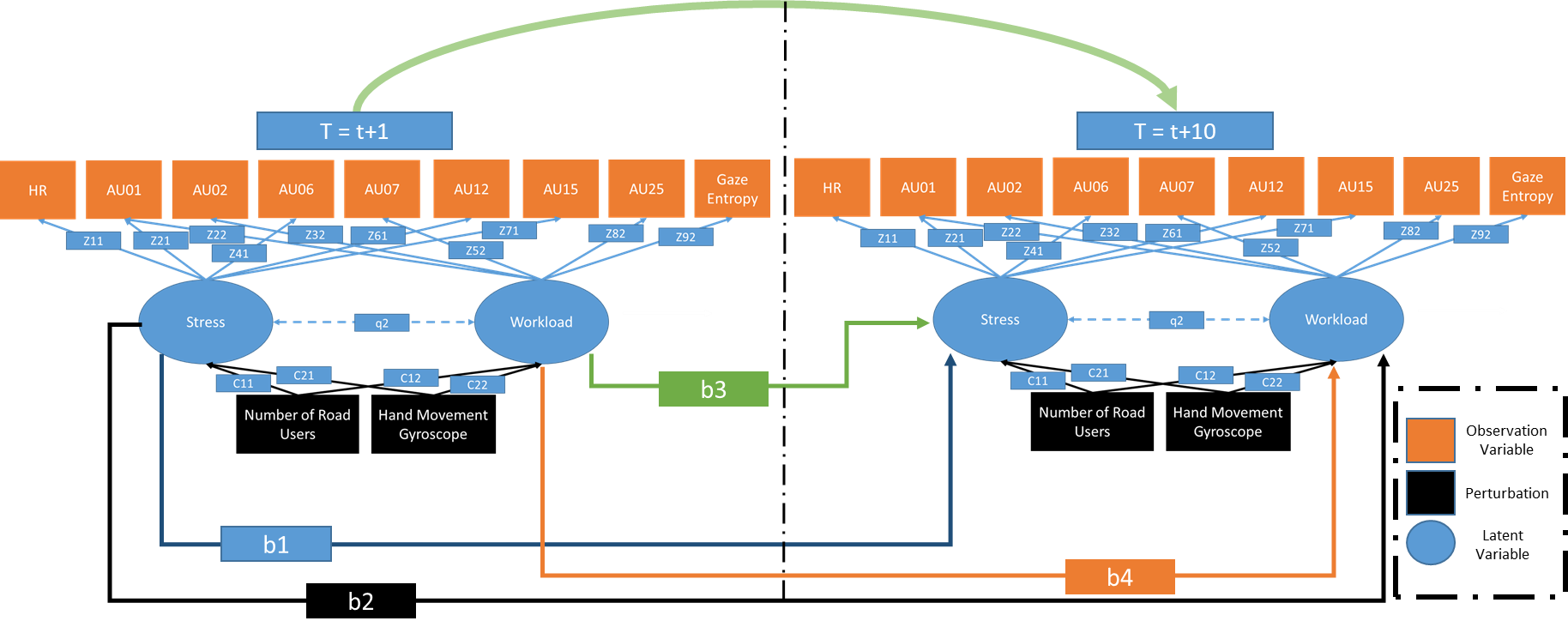}
  \end{center}
  \caption{The two latent variable model. Note that in contrast with this model, in the base model, the two latent variables are identical. In this model a set of perturbations (i.e., number of road users, and hand movement shown with black) is connected to a set of latent variables (i.e., stress level and work load shown with blue) estimated through measurement variables (e.g., HR, AUs, and gaze entropy shown with orange}
  \label{fig:base_mod}
\end{figure*}

\subsection{Model Selection} \label{sec:mod_selection}
In order to answer the research questions, we define two separate models: base model and two latent variables models. In each model, a set of perturbations (i.e., number of road users, and hand movement shown with black on Fig. \ref{fig:base_mod}) is connected to a set of latent variables (i.e., stress level and work load shown with blue on Fig. \ref{fig:base_mod}) estimated through measurement variables (e.g., HR, AUs, and gaze entropy shown with orange on Fig. \ref{fig:base_mod}). 

\textbf{Base Model} This model considers one latent variable referred to as the driver's psychophysiological state. In other words, this model assumes that one latent variable is enough to describe the multimodal data regarding a driver's internal state under the effect of the external context. In order to be able to compare this model with a model with two latent variables, we need to account for the differences in the degrees of freedom. When building this model, we assume two identical latent variables. This helps with accounting for the changes in the degrees of the freedom of the model as compared to the two separate latent variables model, which is described below.

\textbf{Two Latent Variables Model:} The second model includes two separate latent variables for stress and workload of the driver, namely internal context. These latent variables are measured through driver's HR, facial action units, and eye gaze measurement features. This model assumes a covariance between the two latent variables. Thus a covariance matrix between the latent variables is also estimated in the model. More specifically, as shown in Fig. \ref{fig:base_mod}, the q2 variable between stress and workload is estimated through the state space approach. The model is depicted in Fig. \ref{fig:base_mod} for different timesteps. Note that the state-space model takes a recursive attempt at estimating the different matrices as described in section \ref{sec:state_space}. 



\section{Case Study and Results}

In order to choose between the two models, we compare using their -2 $*$ Log Likelihood (-2LL). Table \ref{tab:loglike} shows the -2LL of each model fitted for 11 different participants. As shown, the -2LL of the model with the two latent variable models is considerably lower as compared to the base models. This suggests that the model with interacting two latent variables can better describe the variability in the data for each individual participant. This comparison led us to choose the interacting latent variable model for the rest of our analysis. 

\begin{table}[ht]
\centering
\caption{A comparison between the two models based on their -2LL across different participants. The model with two latent variables has a lower -2LL}
\label{tab:loglike}
\resizebox{0.48\textwidth}{!}{%
\begin{tabular}{cccc}
\textbf{Participant ID} & \textbf{\begin{tabular}[c]{@{}c@{}}Interacting Latent \\ Variable Model\end{tabular}} & \textbf{Base Model} & \textbf{$\Delta$LL} \\ \hline
2  & 170459.8 & 172879   & 2419.2  \\
3  & 299674   & 305882.4 & 6208.4  \\
9  & 272412   & 276434.4 & 4022.4  \\
10 & 101985   & 102780.7 & 795.7   \\
12 & 187456.8 & 189293.8 & 1837    \\
14 & 120738.6 & 122577.1 & 1838.5  \\
16 & 93854.2  & 94914.72 & 1060.52 \\
17 & 153230.4 & 155310.3 & 2079.9  \\
18 & 167789.8 & 170418.8 & 2629    \\
19 & 83357.6  & 84523.7  & 1166.1  \\
20 & 178416.4 & 179560.6 & 1144.2  \\
22 & 208460   & 210691.6 & 2231.6 
\end{tabular}%
}
\end{table}

Let us focus on participant \#9 for a better description of state-space results (Table \ref{tab:estimates})). We then extend our findings to other participants. The latent stress is captured through the higher HR, AU1, AU6, AU15, and lower probability in AU12. The workload is captured through the higher intensity of the AU1, AU2, and lower intensity in AU7, as well as higher gaze entropy. The association between the two latent variables show that higher workload is associated with lower stress level in this participant, which the coefficient is ($Q.q1 = -0.068$ on Table \ref{tab:estimates}). 

\begin{figure}[ht]
\begin{center}
  \frame{\includegraphics[width=1\linewidth]{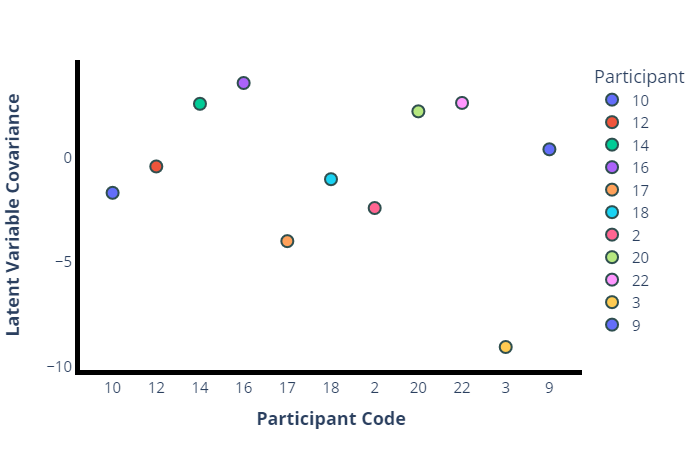}}
  \end{center}
  \caption{The association between the two latent variables of stress and workload across different participants. Note the high individual variability between participants. For some participants the association is positive leading to a synchronized increase in stress level and workload, while for others the two constructs change in different directions.}
  \label{fig:cov_latent}
\end{figure}

The association between the two latent variables varies across participants (Fig. \ref{fig:cov_latent}). Note that these numbers are the result of dividing the association coefficient by the product of the standard deviation of each latent variable. This is required for the correct comparison of the association coefficient across participants. The association coefficient varies between both positive and negative numbers across participants. This can imply that there are two groups of participants. In the first group, an increase in participants' stress level is accompanied by a decrease in workload, while in the second group, the two latent variables change in the same direction. Such high variability across participants shows the importance of individual profiles when considering matters such as distraction, stress levels, and driver state monitoring within different contexts. 

\begin{table}[]
\centering
\caption{Estimates of the interacting latent variable model with their confidence intervals for participant \#9. We specifically focus on one participant to illustrate the results of state space. The parameters shown here are based on the Fig. \ref{fig:base_mod}}
\resizebox{0.4\textwidth}{!}{%
\begin{tabular}{lllll}
 Parameter & ML.Est & Std Error &  low CI & up CI\\\hline
Z11 & 1.76e+00 &0.57898 & 0.62989&  2.899433\\ \hline
Z21 & 2.26e-01 &0.08057 & 0.06782 & 0.383666\\\hline
Z41& -2.50e-02& 0.03421 &-0.09203&  0.042064\\\hline
Z61 & 8.47e-02& 0.04596& -0.00538&  0.174772\\\hline
Z71 & 2.00e-01 &0.07329&  0.05608&  0.343378\\\hline
Z22& -1.87e-05& 0.01258 &-0.02467 & 0.024630\\\hline
Z32 & 2.62e-02& 0.01502& -0.00324&  0.055640\\\hline
Z52& -1.37e-01& 0.04654 &-0.22820 &-0.045754\\\hline
Z82 &-1.21e-01& 0.04164 &-0.20307& -0.039844\\\hline
Z92 & 1.03e+00& 0.33772 & 0.36807 & 1.691904\\\hline
B.b1 &  8.81e-01& 0.00716 & 0.86746 & 0.895518\\\hline
B.b2 &  8.91e-03& 0.00905& -0.00884 & 0.026650\\\hline
B.b3 & -1.58e-03& 0.00271& -0.00689 & 0.003724\\\hline
B.b4 &  9.76e-01& 0.00244 & 0.97149 & 0.981065\\\hline
Q.q2 & -6.08e-03& 0.00422& -0.01435 & 0.002193\\\hline
C.C11 & 9.94e-03 & 0.00398&  0.00215 & 0.017732\\\hline
C.C21 &-6.90e-03 & 0.00326 &-0.01328& -0.000517\\\hline
C.C12& -7.34e-03 &0.00330 &-0.01381 &-0.000870\\\hline
C.C22& -7.53e-03& 0.00339& -0.01418& -0.000877\\\hline
\end{tabular}%
}
\label{tab:estimates}
\end{table}

\textbf{Research Question 1: How do the external context (i.e., traffic density) and task demands (based on hand-movement) affect the driver's internal contextual state (i.e., stress level and workload)?} 

\textbf{Participant \#9:} Let us first start with one participant to illustrate the results. The state-space model for participant 9 shows that the number of road users can be associated with an increase in stress levels and a decrease in workload (compare C.C11 with C.C12 on Table \ref{tab:estimates}). Due to being stuck in higher traffic density with the increase in the number of road users, the mean value of the driver's heart rate increases, which in turn increases the stress level. Previous research has shown that traffic density negatively affects a driver's well-being \cite{morris2016does}. 

Additionally, our model shows that an increase in the number of road users is accompanied by a lower workload (see Table \ref{tab:estimates}). When considering the workload, we have measured the gaze entropy through different regions in the field of view. This method of analysis considers the whole secondary task area (i.e., using the vehicle radio, phone, etc.) as one region. Thus it cannot analyze a person's finer gaze patterns when working with a phone. This is important to consider when analyzing the workload under different environmental perturbations. An explanation for this can be due to the fact that higher traffic density often is accompanied by a stationary state for the vehicle, which motivates the participant to perform secondary tasks more often \cite{young2019contextual}. This decreases the gaze dispersion with respect to the driving scene and diverges it to the secondary task area (e.g., phone and center stack), which decreases the workload as measured with respect to driving and increases the workload spent on the secondary task.  The videos accompanying our analysis were reviewed to confirm this finding qualitatively. Hand movement is associated with a decrease in stress level as well as workload. Note that higher hand movements result in performing secondary tasks in which take the attention from normal driving and bring it into one specific region (i.e., phone and center stack location in the vehicle) (compare C.C21 with C.C22 on Table \ref{tab:estimates}).    

\begin{figure*}[ht]
\begin{center}
  \frame{\includegraphics[width=1\linewidth]{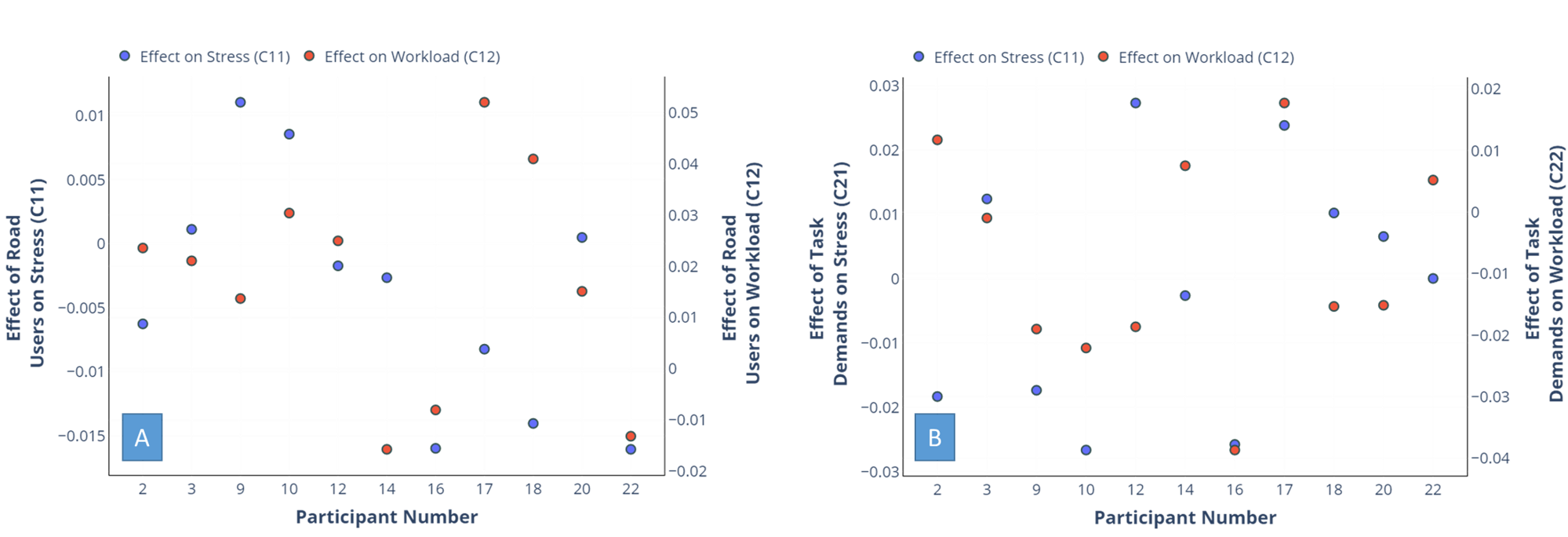}}
  \end{center}
  \caption{The effect of number of road users (A) and task demands (B) on stress and workload across different participants. The number on each marker shows the participant number.}
  \label{fig:road_task}
\end{figure*}

\textbf{Comparison Across Participants:} We then consider comparing different participants with respect to the effect of the external context on their stress level and workload. Fig. \ref{fig:road_task} - A and B depict the effect of number of road users and task demands (i.e., C values on Fig. \ref{fig:base_mod}) on driver's stress and workload respectively. As shown, participants' stress level and work load are associated with different impacts by the number of road users and task demands. While this difference can certainly be due to the different contextual elements that were present in each participant's driving scenario (e.g., weather condition and distance to other vehicles), which we did not account for, it can also be due to individual differences in how they react the number of road users and task demands. In some of the participants, an increase in the number of road users is associated with an increase in their stress level (e.g., participant number 9, 10). For some of the other participants, the increase in the number of road users is associated with a decrease in their stress level (e.g., participants number 16 and 18). There are also some participants that their stress level are not affected by the increase in number of road users (e.g., participant number 3). Simultaneously, the number of road users is associated with an increase in the workload for some participants (e.g., participant numbers 17, and 18) while not for the other participants. Similar results can also be observed in the effect of task demands when comparing different participants (Fig. \ref{fig:road_task} - B). 



While confirming the reason behind the possible individual differences is not possible in a naturalistic study without controlled experiments, one possible explanation can be that an increase in the number of road users might make it harder for some participants to drive, which increases his/her stress level. However, the other group of participants might be using different feedback loops to drain the pressure from the increase in the number of road users, such as using their phones, listening to music, or talking to a passenger. For instance, considering Fig. \ref{fig:road_task} - A, for participant 3, the increase in the number of road users increases their stress while decreasing the workload of driving as measured through gaze entropy. This participant might be using their phone more often in these situations, which is the reason behind the decrease in the level of workload imposed by driving and diverges their attention to the secondary tasks (e.g., phone). The future direction of our research will analyze different feedback loops in participants through differential equation modeling.


\textbf{Research Question 2: How does the latent state at previous timestamps (i.e., 1 through 10 seconds) of the driver affect his/her current state (i.e., stress and workload)?}

\textbf{Participant \#9:} In order to answer this question we focus on the variables connecting the different time steps which are B.b1 and B.b4 on Table \ref{tab:estimates}. It is interesting to observe very high temporal dependency values for drivers' stress and workload. Note that we have already applied a 10-second look ahead for the state-space model. Even with a 10-second look ahead, the model provides evidence that a driver's states (both stress level and workload) are highly dependent on its historical values, meaning that momentary changes can well affect the future state of the driver. 

\begin{figure*}[ht]
\begin{center}
  \frame{\includegraphics[width=1\linewidth]{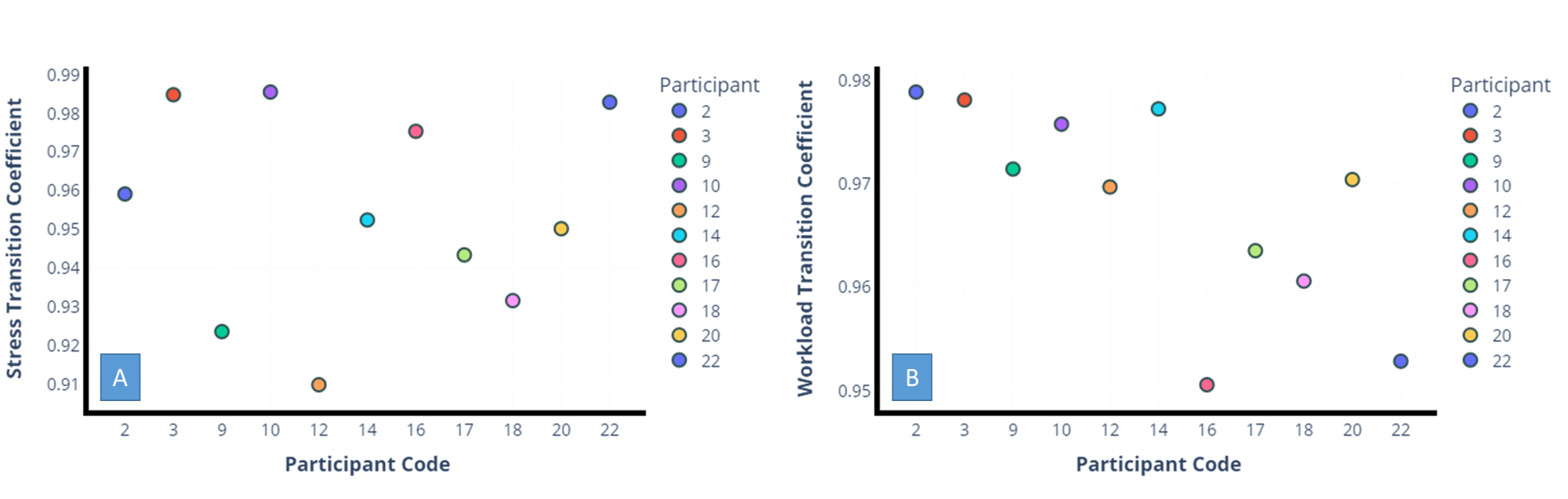}}
  \end{center}
  \caption{The transition coefficient of stress latent variable across participants (A), and the transition coefficient of workload latent variable across participants (B). Note the high dependency of stress latent variable on the previous values across participants}
  \label{fig:stress_workload_trans}
\end{figure*}

\textbf{Comparison Across Participants:} The high dependency is also observed through the data collected from other participants. Fig. \ref{fig:stress_workload_trans} - A and B depict the stress and workload transition coefficients across participants. The high values of coefficients imply that drivers' previous states are well predictive of their future state with a very high dependency when modeled through a state-space framework.  

\begin{figure}[ht]
\begin{center}
  \frame{\includegraphics[width=1\linewidth]{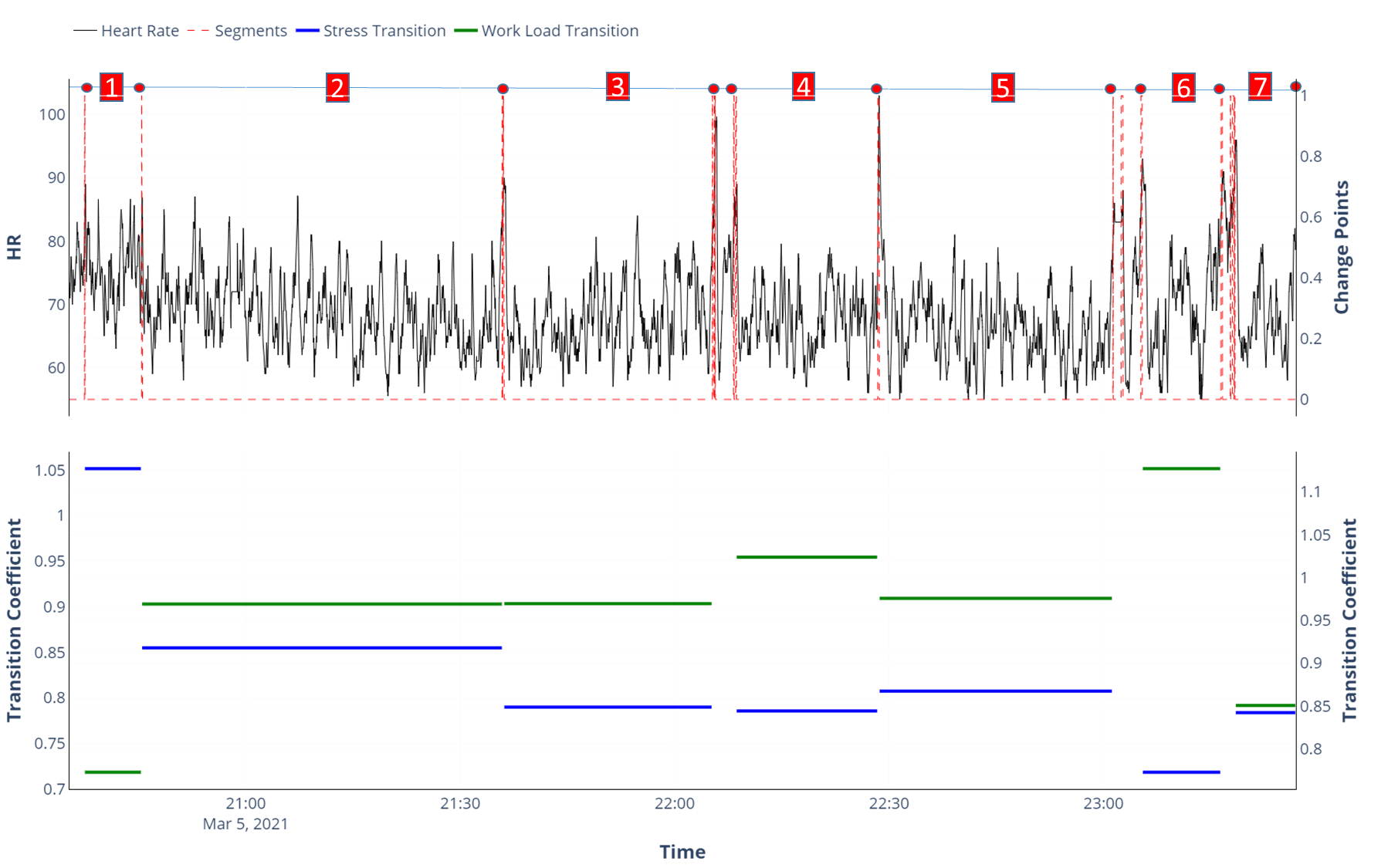}}
  \end{center}
  \caption{Driver's HR together with the locations of segmentation, shown with red lines. These locations are chosen based on the occurrence of a change point with an abrupt increase in HR as much as 1.5 times the standard deviation of HR in the trip.}
  \label{fig:hr_cp}
\end{figure}

One alternative explanation to the high temporal dependency is that the presence of rare events in driving can affect the analysis and may result in inaccurate high dependency values between the driver's states. To detect rare events, we use the results of the change point detection on drivers' HR and find the locations that HR changes abruptly. This analysis is performed based on previous research showing the effect of external context on drivers' HR using changepoint detection methods (for example, see \cite{tavakoli2021harmony,guo2021orclsim,kerautret2022detecting}). To showcase the rare event analysis, let us focus on one participant, \#9. Using the change points in driver's HR, we have segmented the driving scenario for participant 9 to avoid the rare events statistics. Here we define a rare event in driving as an event that is associated with an increase in driver's HR beyond twice the standard deviation of HR in the whole driving scenario. The locations of change points associated with such rare events are shown in Fig. \ref{fig:hr_cp} with vertical red dashed lines. For each segment, we then ran the state-space model. The results of the stress and workload transition coefficients are shown in Fig. \ref{fig:hr_cp}. There are two main observations drawn from Fig. \ref{fig:hr_cp}. Firstly, within different segments as defined by change points, the transition coefficients are still very high (above 0.7), depicting a high dependency on the previous states. Second, the two coefficients do not move in a synchronized fashion. For instance, moving from segment 2 to 3, the workload transition stays constant while stress transition decreases. For this segment, this implies that although the dependency of workload on its historical values does not change, the stress level is becoming more and more unpredictable based upon its historical values. This confirms our finding in the model selection section that this model, in fact, captures two separate constructs through human sensing data, which do not always act in a synchronized fashion. Additionally, this finding lays the ground for possible information transmission between the two constructs as in some cases, they perform synchronized and in other cases, they do not. This brings us to our third research question.

\textbf{Research Question 3: How does the association between latent constructs change throughout the driving scenario?} 

\begin{figure}[ht]
\begin{center}
  \frame{\includegraphics[width=1\linewidth]{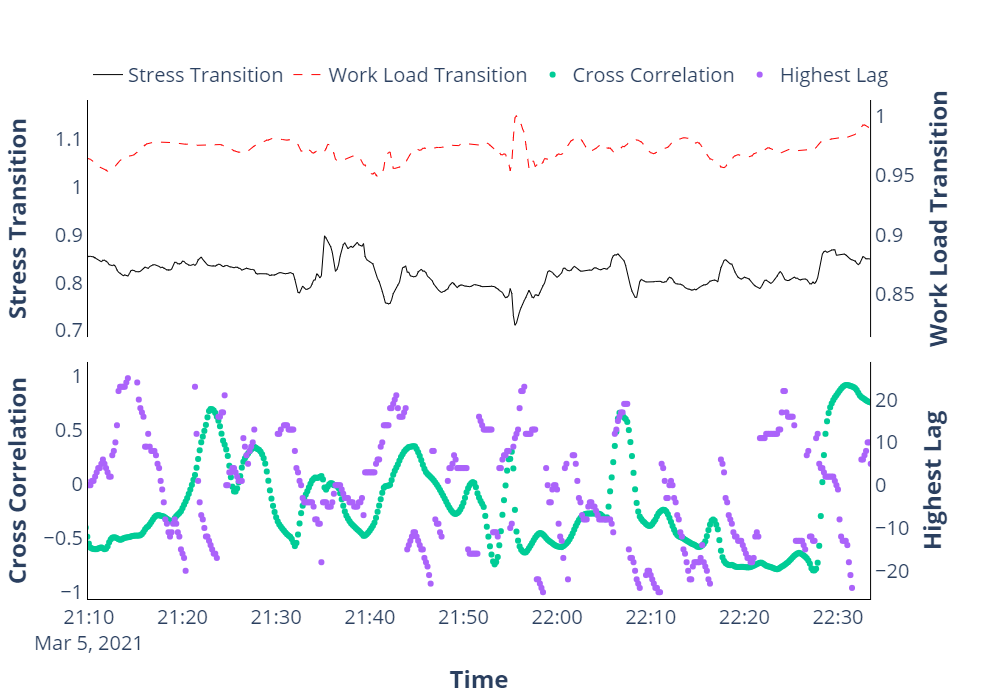}}
  \end{center}
  \caption{The time series of changes in driver's stress and workload transition coefficients. Note that the cross-correlation between the two signals does not always follow the same trend implying that two latent variables are, in fact, measuring two possibly different constructs. Additionally, note that the cross-correlation between the two signals of stress and workload transition coefficient varies during the trip, in which different directions of the flow of information between the two constructs can be seen through the lags across the two signals that maximizes the synchronous behavior.}
  \label{fig:hr_roll}
\end{figure}

To further analyze the association between the two transition coefficients, we have applied a rolling window on the data collected from participant \#9 and assessed a state-space model for each rolling window. More specifically, a sliding window with a length of half an hour of data was considered with an overlap of 1 minute. At each time step, half an hour of data is considered, a state-space model is assessed for the data, and then the window moves for one minute in the forward direction in time, and the analysis is repeated. We have then calculated the cross-correlation between the two signals at each rolling window. This is depicted on Fig. \ref{fig:hr_roll}. The association between the two signals varies across the driving scenario duration, in which it moves between positive to negative values. This might imply that sometimes changes in the two constructs are synchronized, and sometimes they are in opposite direction. 

However, the cross-correlation by itself does not show the exact direction of correlation. In other words, on Fig. \ref{fig:hr_roll} it is not possible to understand whether stress level is leading or following the workload. To further find the answer to the sequence of stress level and workload events, we calculated the windowed cross-correlation and peak picking analysis as provided by \cite{boker2002windowed}. In summary, in order to find out the variation in the association between two behavioral signals, one can perform a windowed cross-correlation (WCC) and peak picking analysis on the data. This analysis technique attempts to find the lag that locally maximizes the correlation between the two signals. This is performed by moving a certain window of each signal in different directions for different lags and assessing the correlation between the two signals. Note that this is different than merely finding the lag that maximizes the correlation, as it performs locally. If the two signals are synchronized, the lag at which the correlation maximizes should always be at zero.

On the other hand, if the dynamics of maximum lag varies during the time, this can imply that the sequence of the two signals can change throughout time. In other words, in our case, it can imply that sometimes events in stress level proceed the events in workload and sometimes follows. Although this does not necessarily suggest evidence on causality, as both latent variable events can be caused by a third variable, one possible explanation in such situations can be that stress causes workload at some time points, and in the other ones, workload causes stress. Another possible explanation is that the response in stress level and workload to different inputs happen at different time scales in which sometimes the stress level responds faster and in other times the workload leads the response.

After performing the WCC analysis on the workload and stress transition coefficients, the results are plotted on Fig. \ref{fig:hr_roll} as purple dots showing the lag at which the correlation at each window maximizes. There are two main observations in this plot that helps with understanding the relationship between stress level and workload transition coefficients. Firstly, the lag between the two signals moves from positive to negative values showing changes in the sequence of the two signals of stress and workload transition. This might imply that there is a flow of information between the two constructs that move in different directions at different time points. Second, it is also visible that as the direction starts to change, the correlation between the two signals calculated through cross-correlation starts to increase. 

\section{Discussion}
In this study, we have analyzed the interaction between the driver's state (i.e., internal context) and the environmental attributes (i.e., external context) through modeling the system of the driver and the environment using a state-space fashion. Using a state-space modeling approach, we demonstrated that a two-latent variable model could better describe the multimodal driver psychophysiological sensing data, pointing out the possibility of having multiple psychological constructs that interact with one another. Additionally, our model takes a holistic approach in analyzing the relationship between the external and internal context by analyzing the state of the driver in the environment through state-space latent variable modeling. 

In our study, we observed a strong dependency of the latent variables on their values from 10 seconds prior. This has strong implications for designing autonomous vehicles where drivers' state is sought to be estimated and predicted for a safer shared autonomy. Current guidelines of National Highway Traffic Safety Administration \cite{nhtsa}, define different levels of automation from 0 to 5, where in level 3 of automation, the driver is not required to monitor the road but has to take-over the control with prior notice. Our study has implications for defining a proper notice period for take-over control while considering drivers' state. While we have focused on a 10 second look-ahead time window, more studies are required to find the dependency between drivers' state, and its historical values for a bigger population and within different contexts (e.g., city versus highway). When considering the effect of emotions on a driver's take-over control, it is important to consider the autoregressive nature of different driver state constructs the fact that prior states such as stress level are associated with the current state. Additionally, when analyzing prior naturalistic driving data for gaining insights regarding the effect of environmental factors on safety and crash-related matters, the analysis should move beyond a couple of seconds prior to events. For example, when considering the effect of driver's emotions or secondary task engagements on the prevalence of accidents (e.g., for a sample prevalence analysis, see \cite{dingus2016driver}), it is important to move beyond a couple of seconds prior to the accident, as drivers' state is highly associated to its previous values. Thus a stressful event can well propagate throughout the future timesteps.

Driving is a cognitive task that can often be accompanied by acute stress. Certain on-road (e.g., getting closer to a lead vehicle), and internal characteristics of the driver (e.g., remembering a stressful event) can all affect the driver's state simultaneously. In our study, we observe a better model fit for two latent variables as compared to one when considering driver's state measures such as HR, facial, and gaze. A two latent variable model provides a better description for the underlying variability in the driver's state measures. Considering previous studies pointing to a need for modeling schemes that integrate workload and affective states together \cite{jeon2015towards}, this might provide evidence for the existence of different constructs when analyzing driver state data. This implies that driver state modeling should take a holistic approach and consider multiple constructs rather than isolating each construct. Thus a naturalistic study focused on emotions in driving may need to also consider the workload and vice versa. 

The association between stress level and workload varies throughout the study. While we did not analyze the reason behind variation, there seems to be a flow of information between different constructs that, in some cases, one leads the other one. Although this might not provide full evidence for a causal relationship between stress level and workload, it might imply that the response to the environment happens with different lags within the two constructs. The value and also the direction of the lag are not constant and vary in different driving situations. This implies the importance of considering context when analyzing psychological constructs in a holistic approach. 

The state-space approach estimates the latent variables through different observatory variables; thus, any new measurement can also be added to the modeling scheme as they become available. This emphasizes the modular nature of the state-space approach in estimating the latent variable and the effect of the perturbations upon them. For instance, different sensors are being added to the wearable devices continuously, such as skin temperature, skin conductance, etc. Every new sensor can be added as a measurement variable. Similarly, environmental measurements such as distance to other vehicles, road complexity measures, etc., can also be added as perturbations to the system in the current modeling scheme.


Lastly, recent psychological theories and experimental findings are providing more and more evidence on the importance of individualizing profiles for driver state analysis. In other words, without considering individual differences, building generalizable frameworks for driver state detection might easily fail when applying them in subject-independent situations. In our study, we observe that even within similar environments, the association between participants' psychological constructs can be very different across our participants. Additionally, we observe that external contextual elements such as traffic density can have different effects on different participants. In some participants, it can increase the stress level, while in other participants, it might decrease the stress level. It is important to note that higher traffic density, while being stressful for many individuals, can actually decrease the speed of travel, which might be pleasant to some other participants. This might explain the differences across participants in the coefficients retrieved through state-space models. More studies are required to analyze the relationship between the driver and the environment within varying contexts.  

\section{Limitations}

There are currently a set of limitations that will be addressed in our future work. We will expand the number of participants to understand the individual differences across different scenarios. For instance, for our study, we have focused on one highway driving scenario for each participant, whereas other scenarios might provide additional information on the individual differences in the interaction between the two constructs. While in this study, we did not find meaningful differences across age and gender, future work will also consider the variation in the impact of the environment with respect to different ages and gender. 

Additionally, we will increase the number of features to include distance to the other vehicles, lane position, and speed that are detected from the external environment. Adding more modalities might provide information on the differences across participants on the impact of environment on their stress and workload. For example, it could be the case that two drivers are facing a similar number of road users but with different distances to the lead vehicles. Additionally, this will help us better understand the interaction between the contextual elements. Using other behavioral metrics such as driver's speed, we can better analyze the driver's feedback to the elements of external context. The feedback loop will be analyzed in greater detail using second-order differential equation models where we account explicitly for a participant's resilience in different driving events.

\section{Conclusion}
In this research, we propose using the latent variable state-space modeling approach for analyzing the impact of the external context on drivers' states. Through applying this modeling scheme on naturalistic driving data retrieved through HARMONY, (1) we estimate driver's stress level and workload by using the data from driver's cardiovascular measures as well as gaze variability and facial expression data; (2) we estimate the effect of the number of cars and drivers task demands as perturbations to this dynamical system on driver's stress level and workload; and (3) we analyze the temporal dependency of driver's state during a driving scenario. Our work paves the way for designing human-centered driver-vehicle interaction systems that can understand and respond to changes in the driver's state resulting from the driving environment and provide a safer driving experience.

\bibliographystyle{IEEEtran}
\bibliography{bib}

\begin{IEEEbiography}[{\includegraphics[width=1in,height=1.25in,clip,keepaspectratio]{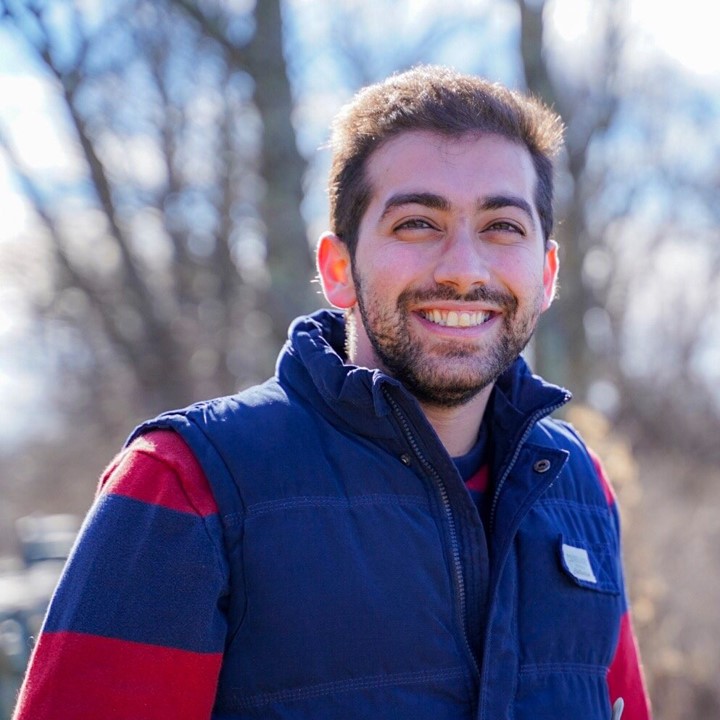}}]{Arash Tavakoli} Arash Tavakoli is a Ph.D. candidate in the Engineering Systems and Environment department as well as the Link Lab at the University of Virginia. He has earned his BSc. and MSc. in Civil Engineering from the Sharif University of Technology and Virginia Tech, respectively. Arash’s research interest lies on the intersection of transportation engineering, computer science, and psychology.
\end{IEEEbiography}

\begin{IEEEbiography}[{\includegraphics[width=1in,height=1.25in,clip,keepaspectratio]{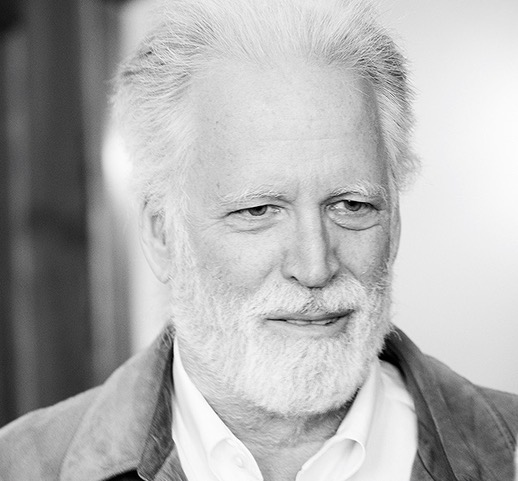}}]{Steven Boker} Professor Steven M. Boker, Ph.D., directs the quantitative psychology program at the University of Virginia.  His contributions include Latent Differential Equations, Generalized Local Linear Approximation, and Windowed Cross-Correlation for dynamical systems modeling of time series.  He is also a lead software architect of the OpenMx Structural Equation Modeling program.

\end{IEEEbiography}

\begin{IEEEbiography}[{\includegraphics[width=1in,height=1.25in,clip,keepaspectratio]{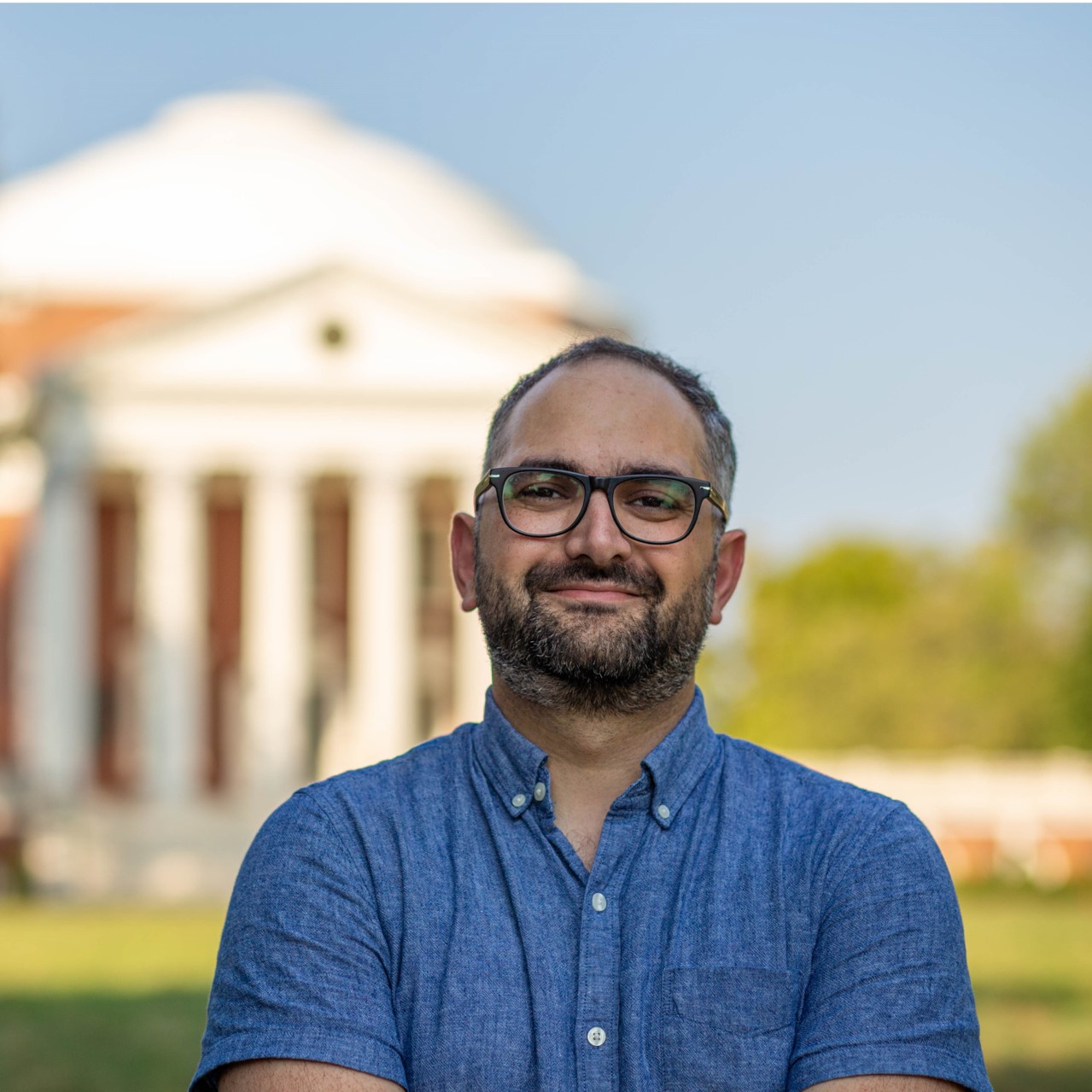}}]{Arsalan Heydarian} Dr. Arsalan Heydarian is an Assistant Professor in the Department of Engineering Systems and Environment as well as the UVA LINK LAB. His research focuses on user-centered design, construction, and operation of intelligent infrastructure with the objective of enhancing sustainability, adaptability, and resilience in future infrastructure systems. Specifically, his research can be divided into four main research streams: (1) intelligent built environments; (2) mobility and infrastructure design; (3) smart transportation; and (4) data-driven mixed reality. Dr. Heydarian received his Ph.D. in Civil Engineering from the University of Southern California (USC), M.Sc in System Engineering from USC, and B.Sc. and M.Sc in Civil Engineering from Virginia Tech. 
\end{IEEEbiography}

\end{document}